\documentclass[12pt,aps,prd,showpacs,amsmath,amssymb]{revtex4}
\input epsf
\begin{document}
\title{Higher order light-cone distribution amplitudes of the Lambda baryon}
\author{Yong-Lu Liu$^1$, Chun-Yu Cui$^2$, and Ming-Qiu Huang$^1$}
\affiliation{$^1$ College of Science, National University of Defense Technology, Hunan 410073, China}
\affiliation{$^2$ Department of Physics, School of Biomedical Engineering, Third Military Medical University, Chongqing 400038, China}
\date{\today}
\begin{abstract}
The improved light-cone distribution amplitudes (LCDAs) of the $\Lambda$ baryon are examined on the basis of the QCD conformal partial wave expansion approach. The calculations are carried out to the next-to-leading order of conformal spin accuracy with consideration of twist $6$. The next leading order conformal expansion coefficients are related to the nonperturbative parameters defined by the local three quark operator matrix elements with different Lorentz structures with a covariant derivative. The nonperturbative parameters are determined with the QCD sum rule method. The explicit expressions of the LCDAs are provided as the main results.
\end{abstract}
\pacs{11.25.Hf, 11.55.Hx, 14.20.Dh.} \maketitle

\section{Introduction}\label{sec1}

Many achievements have been made in the past years in the area of the elementary particle physics at high energy scale with the experiments carried out at LHC, BESIII, RIHC and other high energy factories. However, we have to conquer the nonperturbative difficulties in QCD whenever dealing with phenomena in hadron physics. The nonperturbative effect is related to the intrinsic structure of the QCD vacuum. Before this problem is resolved completely, some effective nonperturbative tools are still needed in order to analyze physics related to the hadronic scale. QCD sum rules is a useful tool to estimate unknown hadronic parameters. Calculating the correlation functions that involve hadron properties both at quark level and hadronic level, the hadronic parameters can be estimated by matching the phenomenon and theoretical sides with quark-hadron duality approach \cite{svzsum}. Light-cone QCD sum rules (LCSR)\cite{lcsr1,lcsr2,lcsr3} is a development of the traditional QCD sum rules. The fundamental inputs in LCSR are the light-cone distribution amplitudes, which describe the distribution of the hadron momentum on the particles inside the hadron.

Light-cone QCD sum rules has been adopted to study properties of baryon physics, especially the heavy flavor physics\cite{heavydecay} and electromagnetic form factors\cite{elemagff}, after their successful applications of the heavy to light meson transition form factors\cite{heavymeson}. Some of the previous works \cite{dynamic} shows that when using LCDAs from conformal spin expansion approach, the next-to-leading order corrections may affect the results to some extent, particularly for the dynamical processes. In recent years, much data on the heavy $\Lambda_b$ baryon has been accumulated at LHC\cite{lamb} and CDF\cite{lamb2}, which provide an excellent platform to investigate the intrinsic properties of QCD. Therefore the higher order corrections to the LCDAs of the $\Lambda$ baryon is instructive in investigating the nonperturbative effects at hadron level.

In the previous works, we present the light-cone distribution amplitudes of $\Sigma$ baryon\cite{DAs-sig} and the higher order corrections to the next-to-leading order accuracy of the conformal spin expansion\cite{DAs-sighigherorder}. This article is a complement to the previous ones and we aim to present the explicit expressions of the light-cone distribution amplitudes of the $\Lambda$ baryon in the approach of conformal spin expansion\cite{Braun1,Braun2,Lenz,Balitsky} to the next-to-leading order. In comparison with the nucleon or the $\Sigma$ and $\Xi$ baryon\cite{DAs-sig,DAs-xi}, $\Lambda$ has isospin $0$, which makes it loose the symmetry relationships from the identity of the quarks. As a result, the equations to reduce the freedom of the nonperturbative parameters have different origins from that of the other Octet $J^P=\frac{1}{2}^+$ baryons. One of the main results of the paper is to give such relations.

It is the same as that in Ref. \cite{DAs-sighigherorder}, we do not consider higher twist effects from four-particle contributions. The higher order conformal spin contribution comes from the higher expansion of the non-local three-quark matrix element between vacuum and the baryon state. The nonperturbative parameters are determined by the higher moment of the local composite operators of the baryon, which can be defined by different coupling constants. Then these coupling constants are connected to the parameters of the conformal spin expansion.

The rest of the paper is organized as follows. Section \ref{sec:def} is devoted to present a general frame work of the LCDAs. The conformal partial wave expansion of the LCDAs is carried out by use of the conformal symmetry of the massless QCD Lagrangian. The equations of motion are used to reduce the number of the free parameters from $24$ to $10$. The nonperturbative parameters connected with the LCDAs are determined in Sec. \ref{sec:sumrule} with the QCD sum rule method. Finally, we give the explicit expressions of the $\Lambda$ baryon LCDAs in Sec. \ref{sec:result}. Section \ref{sec:summary} is a summary of the paper.

\section{Definitions and conformal expansion}\label{sec:def}

Light-cone distribution amplitudes of hadrons characterize the nonperturbative properties of the hadrons by describing the momentum distribution of the partons inside the composite particle. LCDAs are fundamental parameters in light-cone QCD sum rules and the hard exclusive processes theory \cite{exclusive,exclusive2}. It is the same as definitions in Refs. \cite{Chernyak,Braun1,DAs-sig}, the LCDAs of the $\Lambda$  baryon can be defined by the general Lorentz expansion of the matrix element of the nonlocal three-quark-operator between the vacuum and the baryon state
\begin{equation}
\langle{0} |\epsilon^{ijk} u_\alpha^i(a_1 z) d_\beta^j(a_2 z) s_\gamma^k(a_3 z) |{\Lambda(P)}\rangle \,,\label{matele}
\end{equation}
where $\alpha, \beta, \gamma$ are Lorentz indices and $i, j, k$ represent color ones. It is pointed out that the gauge factor $[x,y]=P\mbox{exp}[ig_s\int_0^1dt(x-y)_\mu A^\mu(tx+(1-t)y)]$ need to be inserted to make the matrix element above gauge invariant. In practice, fixed-point gauge $(x-y)^\mu A_\mu(x-y)=0$ is adopted so that this factor is equal to unity. Thus in this paper we do not show them explicitly.

The matrix element (\ref{matele}) can be generally decomposed in consideration of the Lorentz covariance, spin and parity properties of the $\Lambda$ baryon as follows:
\begin{equation}
4 \langle{0} |\epsilon^{ijk} u_\alpha^i(a_1 z) d_\beta^j(a_2 z) s_\gamma^k(a_3 z) |{\Lambda(P)}\rangle = \sum\limits_{i}\mathcal{F}_i\,\Gamma_{1i}^{\alpha\beta}\Big
(\Gamma_{2i}\Lambda\Big )_\gamma \,,\label{da-def}
\end{equation}
where $\Lambda_\gamma$ is the spinor of the baryon with the quantum number $I(J^P)=0(\frac{1}{2}^+)$ ($I$ is the isospin, $J$ is the total angular momentum, and
$P$ is the parity), $\Gamma_{1(2)i}$ are certain Dirac structures over which the sum is carried out, and $\mathcal{F}_i=\mathcal{S}_i, \mathcal {P}_i, \mathcal
{A}_i, \mathcal{V}_i,\mathcal{T}_i$ are the independent distribution amplitudes which are functions of the scalar product $P\cdot z$\cite{DAs-sig}. It is also noticed that $z$ and $p$ are vectors defined on the light-cone: $z^2=0$ and $p^2=0$.

The matrix element in Eq. (\ref{da-def}) can also be decomposed with definite twist and correspondingly the LCDAs are defined as $F_i$ in the infinite momentum frame as:
\begin{equation}
4\langle {0}| \epsilon^{ijk} u_\alpha^i(a_1 z) d_\beta^j(a_2 z) s_\gamma^k(a_3 z) |{\Lambda(P)}\rangle =\sum\limits_{i}F_i\,\Gamma_{1i}'^{\alpha\beta}\Big
(\Gamma_{2i}'\Lambda\Big )_\gamma\,.\label{da-deftwist}
\end{equation}
These two sets of definitions have the following relations:
\begin{eqnarray}
&{\cal S}_1 = S_1\,, & 2p\cdot z\,{\cal S}_2 = S_1-S_2\,, \nonumber\\
&{\cal P}_1 = P_1\,, & 2p\cdot z\,{\cal P}_2 = P_2-P_1\,,\nonumber\\
&\mathcal V_1 = V_1\,, & 2 p\cdot z\mathcal V_2 = V_1 - V_2 - V_3\,,\nonumber \\
&2 \mathcal V_3 = V_3\,, & 4 p\cdot z\mathcal V_4 = - 2 V_1 + V_3 + V_4 + 2 V_5\,,\nonumber \\
&4 p\cdot z \mathcal V_5 = V_4 - V_3\,, & (2 p\cdot z )^2\mathcal V_6 = - V_1 + V_2 + V_3 + V_4 + V_5 - V_6
\end{eqnarray}
for scalar, pseudoscalar, and vector structures, and
\begin{eqnarray}
&\mathcal A_1 = A_1\,, & 2 p\cdot z\mathcal A_2 = - A_1 + A_2 - A_3\,, \nonumber\\
&2 \mathcal A_3 = A_3\,, & 4 p\cdot z\mathcal A_4 = - 2 A_1 - A_3 - A_4 + 2 A_5\,, \nonumber\\
&4 p\cdot z \mathcal A_5 = A_3 - A_4\,, & (2 p\cdot z )^2\mathcal A_6 = A_1 - A_2 + A_3 + A_4 - A_5 + A_6
\end{eqnarray}
for axial-vector structures, and
\begin{eqnarray}
&\mathcal T_1 = T_1\,, & 2 p\cdot z\mathcal T_2 = T_1 + T_2 - 2 T_3\,, \nonumber\\
&2 \mathcal T_3 = T_7\,,& 2 p\cdot z\mathcal T_4 = T_1 - T_2 - 2 T_7\,, \nonumber\\
&2 p\cdot z \mathcal T_5 = - T_1 + T_5 + 2 T_8\,, &
(2 p\cdot z)^2\mathcal T_6 = 2 T_2 - 2 T_3 - 2 T_4 + 2 T_5 + 2 T_7 + 2 T_8\,, \nonumber\\
&4 p\cdot z \mathcal T_7 = T_7 - T_8\,, &(2 p\cdot z)^2\mathcal T_8 = -T_1 + T_2 + T_5 - T_6 + 2 T_7 + 2 T_8
\end{eqnarray}
for tensor structures.

The classifications of the LCDAs $F_i$ with a definite twist and the explicit expressions of the definition can be found in Refs. \cite{Braun1,DAs-sig}. Each distribution amplitude $F_i$ can be represented as
\begin{equation}
F(a_ip\cdot z)=\int \mathcal Dxe^{-ipz\sum\limits_ix_ia_i}F(x_i),
\end{equation}
where the dimensionless variables $x_i$ meet the relations $0<x_i<1$ and $\sum\limits_ix_i=1$, corresponding to the longitudinal momentum fractions along the light-cone carried by the quarks inside the baryon. The integration measure is defined as
\begin{equation}
\int \mathcal Dx=\int_0^1dx_1dx_2dx_3\delta(x_1+x_2+x_3-1).
\end{equation}

The $\Lambda$ baryon has the isospin $I=0$, which is useful to deduce the following relationship
\begin{equation}
\langle{0} |\epsilon^{ijk} u_\alpha^i(1) d_\beta^j(2) s_\gamma^k(3) |{\Lambda(P)}\rangle=-\langle{0} |\epsilon^{ijk} u_\beta^i(2) d_\alpha^j(1) s_\gamma^k(3) |{\Lambda(P)}\rangle.\label{isospin}
\end{equation}
This relation can be used to reduce the number of the independent functions. Taking into account the Lorentz decomposition of the $\gamma$-matrix structure, it is easy to see that the vector and tensor LCDAs are antisymmetric under the exchange of the $u$ and $d$ quarks, whereas the scalar, pseudoscalar and axial-vector structures are symmetric:
\begin{eqnarray}
V_i(1,2,3)&=&-V_i(2,1,3),\hspace{2.0cm} T_i(1,2,3)=-T_i(2,1,3),\nonumber\\
S_i(1,2,3)&=&S_i(2,1,3),\hspace{2.3cm} P_i(1,2,3)=P(2,1,3),\nonumber\\
A_i(1,2,3)&=&A(2,1,3).\label{symm-iso}
\end{eqnarray}
The ``calligraphic'' structures in Eq. (\ref{da-def}) have the similar relationships.

In order to get the LCDAs of the baryon, we need express the LCDAs defined above with chiral field representation, so as to use the conformal symmetry of the massless QCD Lagrangian. The explicit expressions and the relations between different definitions are referred to Ref. \cite{DAs-sig}. Here after we only present the conformal expansions according to their twists:

\begin{eqnarray}
\Phi_3(x_i)&=&120x_1x_2x_3[\phi_3^0+\phi_3^-(x_1-x_2)+\phi_3^+(1-3x_3)+...],\nonumber\\
 T_1(x_i)&=&120x_1x_2x_3[t_1^0+t_1^-(x_1-x_2)+t_1^+(1-3x_3)+...]\label{contwist3}
\end{eqnarray}
for twist-$3$ and
\begin{eqnarray}
\Phi_4(x_i)&=&24x_1x_2[\phi_4^0+\phi_4^-(x_1-x_2)+\phi_4^+(1-5x_3)+...],\nonumber\\
\Psi_4(x_i)&=&24x_1x_3[\psi_4^0+\psi_4^-(x_1-x_3)+\psi_4^+(1-5x_2)+...],\nonumber\\
\Xi_4(x_i)&=&24x_2x_3[\xi_4^0+\xi_4^-(x_2-x_3)+\xi_4^+(1-5x_1)+...],\nonumber\\
{\Xi'}_4(x_i)&=&24x_2x_3[{\xi'}_4^0+{\xi'}_4^-(x_2-x_3)+{\xi'}_4^+(1-5x_1)+...],\nonumber\\
T_2(x_i)&=&24x_1x_2[t_2^0+t_2^-(x_1-x_2)+t_2^+(1-5x_3)+...]\label{contwist4}
\end{eqnarray}
for twist-$4$ and
\begin{eqnarray}
\Phi_5(x_i)&=&6x_3[\phi_5^0+\phi_5^-(x_1-x_2)+\phi_5^+(1-2x_3)+...],\nonumber\\
\Psi_5(x_i)&=&6x_2[\psi_5^0+\psi_5^-(x_1-x_3)+\psi_5^+(1-2x_2)+...],\nonumber\\
\Xi_5(x_i)&=&6x_1[\xi_5^0+\xi_5^-(x_2-x_3)+\xi_5^+(1-2x_1)+...],\nonumber\\
{\Xi'}_5(x_i)&=&6x_1[{\xi'}_5^0+{\xi'}_5^-(x_2-x_3)+{\xi'}_5^+(1-2x_1)+...],\nonumber\\
T_5(x_i)&=&6x_3[t_5^0+t_5^-(x_1-x_2)+t_5^+(1-2x_3)+...]\label{contwist5}
\end{eqnarray}
for twist-$5$, and
\begin{eqnarray}
\Phi_6(x_i)&=&2[\phi_6^0+\phi_6^-(x_1-x_2)+\phi_6^+(1-3x_3)+...],\nonumber\\
 T_6(x_i)&=&2[t_6^0+t_6^-(x_1-x_2)+t_6^+(1-3x_3)+...] \label{contwist6}
\end{eqnarray}
for twist-$6$. There are altogether $42$ expansion coefficients which need to be determined (In fact, the free parameters in functions $T_i(i=1,2,5,6)$ can be reduced due to the symmetry relationships in Eq. (\ref{symm-iso}), which will be given at the end of this section).

To the next-to-leading order, the normalization of the $\Lambda$ baryon LCDAs is determined by the matrix element of the nonlocal three-quark operator expanded to the next leading order at the zero point. The expansion of the matrix element is
\begin{eqnarray}
\langle0|\epsilon^{ijk}u^i_\alpha(a_1z)d^j_\beta(a_2z)s^k_\gamma(a_3z)|\Lambda(P)\rangle
=\langle0|\epsilon^{ijk}u^i_\alpha(0)d^j_\beta(0)s^k_\gamma(0)|\Lambda(P)\rangle\nonumber\\
+z_\lambda\langle0|[\epsilon^{ijk}u^i_\alpha(a_1z)\stackrel{\leftrightarrow}{D}d^j_\beta(a_2z)]s^k_\gamma(a_3z)|\Lambda(P)\rangle|_{z=0}\nonumber\\
+z_\lambda\langle0|\epsilon^{ijk}u^i_\alpha(a_1z)d^j_\beta(a_2z)[\vec Ds^k_\gamma(a_3z)]|\Lambda(P)\rangle|_{z=0}.\label{localexp}
\end{eqnarray}

The local operator matrix element on the right side of Eq. (\ref{localexp}) can be generally decomposed according to the Lorentz structure as
\begin{eqnarray}
4\langle0|\epsilon^{ijk}u^i_\alpha(0)d^j_\beta(0)s^k_\gamma(0)|\Lambda(P)\rangle=\mathcal{S}^0_1MC_{\alpha\beta}(\gamma_5\Lambda)_\gamma+\mathcal{P}^0_1M(\gamma_5
C)_{\alpha\beta}\Lambda_\gamma\nonumber\\
+\mathcal{A}^0_1(\!\not\!
P\gamma_5C)_{\alpha \beta}\Lambda_\gamma+\mathcal{A}^0_3M(\gamma_\mu \gamma_5C)_{\alpha
\beta}(\gamma_\mu\Lambda)_\gamma
\end{eqnarray}
for the matrix element of the leading order, and
\begin{eqnarray}
&&4\langle0|\epsilon^{ijk}u^i_\alpha(a_1z)d^j_\beta(a_2z)[\vec D_\lambda s^k_\gamma(a_3z)]|\Lambda(P)\rangle|_{z=0}\nonumber\\
&&=\mathcal{S}_1^sMP_\lambda C_{\alpha\beta}(\gamma_5\Lambda)_\gamma+\mathcal{S}_2^0M^2C_{\alpha\beta}(\gamma_\lambda\gamma_5\Lambda)_\gamma+\mathcal{P}_1^sP_\lambda M(\gamma_5
C)_{\alpha\beta}\Lambda_\gamma\nonumber\\
&&+\mathcal{P}_2^0M^2(\gamma_5 C)_{\alpha\beta}(\gamma_\lambda\Lambda)_\gamma
+\mathcal{A}_1^sP_\Lambda(\!\not\!
P\gamma_5C)_{\alpha\beta}\Lambda_\gamma
+\mathcal{A}_2^0M(\!\not\!
P\gamma_5C)_{\alpha \beta}(\gamma_5\Lambda)_\gamma\nonumber\\
&&+\mathcal{A}_3^sP_\lambda M(\gamma_\mu\gamma_5C){\gamma_\mu\Lambda}_\gamma+\mathcal{A}_4^0M^2(\gamma_\lambda\gamma_5C)_{\alpha
\beta}(\sigma^{\mu\nu}\gamma_5\Sigma)_\gamma+\mathcal{A}_5^0M^2(\gamma_\mu\gamma_5C)_{\alpha
\beta}(i\sigma^{\mu\lambda}\Lambda)_\gamma,\nonumber\\
\end{eqnarray}
\begin{eqnarray}
&&4\langle0|[\epsilon^{ijk}u^i_\alpha(a_1z)\stackrel{\leftrightarrow}{D}d^j_\beta(a_2z)]s^k_\gamma(a_3z)|\Lambda(P)\rangle|_{z=0}\nonumber\\
&&=\mathcal{V}_1^0P_\lambda (\!\not\!
PC)_{\alpha\beta}(\gamma_5\Lambda)_\gamma
+\mathcal{V}_2^0M(\!\not\!
PC)_{\alpha\beta}(\gamma_\lambda\gamma_5\Lambda)_\gamma+\mathcal{V}_3^0P_\lambda M(\gamma_\mu
C)_{\alpha\beta}{\gamma_\mu\gamma_5\Lambda}_\gamma\nonumber\\
&&+\mathcal{V}_4^0M^2(\gamma_\lambda C)_{\alpha\beta}(\gamma_5\Lambda)_\gamma+\mathcal{V}_5^0M^2(\gamma_\mu C)_{\alpha\beta}{i\sigma{\mu\lambda}\gamma_5}_\gamma+\mathcal{T}_1^0P_\lambda(P\nu i\sigma_{\mu\nu}C)_{\alpha\beta}{\gamma_\mu\gamma_5\Lambda}_\gamma\nonumber\\
&&+\mathcal{T}_2^0 M(P_\mu i\sigma_{\lambda\nu} C)_{\alpha\beta}{\gamma_5\Lambda}_\gamma+\mathcal{T}_3^0P_\lambda M(\sigma_{\mu\nu} C)_{\alpha\beta}(\sigma^{\mu\nu}\gamma_5\Lambda)_\gamma+\mathcal{T}_4^0 M(P^\mu\sigma_{\mu\nu} C)_{\alpha\beta}{\sigma_{\mu\lambda}\gamma_5\Lambda}_\gamma\nonumber\\
&&+\mathcal{T}_5^0M^2(i\sigma_{\mu\lambda}C)_{\alpha\beta}{\gamma_\mu\gamma_5\Lambda}_\gamma+\mathcal{T}_7^0P_\lambda M^2(\sigma_{\mu\nu} C)_{\alpha\beta}(\sigma^{\mu\nu}\gamma_\lambda\gamma_5\Lambda)_\gamma
\end{eqnarray}
for the next leading order ones. There are altogether $24$ nonperturbative parameters in the expressions. However, the parameters defined above are not independent and can be reduced with the help of the equations of motion. The constraints are:
\begin{eqnarray}
&&
\langle{0}| \epsilon^{ijk} u^i(0) C \gamma_5\gamma_\rho d^j(0)\gamma^\lambda [iD_\lambda s_\gamma]^k(0)|{\Lambda,P} \rangle = 0 \,,
\nonumber \\
&&
\langle{0}| \epsilon^{ijk} u^i(0) C \gamma_5\gamma^\lambda d^j(0)
[iD_\lambda s_\gamma]^k(0)|{\Lambda,P}\rangle =
P_\lambda
\langle{0}| \epsilon^{ijk} u^i(0) C \gamma_5\gamma_\lambda d^j(0)
s_\gamma^k(0)|{\Lambda,P}\rangle \,,
\nonumber \\
&&
\langle{0}| \epsilon^{ijk} u^i(0) C \gamma_5 d^j(0)
[iD_\lambda s_\gamma]^k(0)|{\Lambda,P}\rangle\nonumber\\
&& = P_\lambda\langle{0}|\epsilon^{ijk}u^i(0)c\gamma_5d^j(0)s_\gamma^k(0)|{\Lambda,P}\rangle-\langle{0}|\epsilon^{ijk}u^i(0)\stackrel{\leftrightarrow}{i D^\lambda}c\gamma_5d^j(0)s_\gamma^k(0)|{\Lambda,P}\rangle \,,
\nonumber \\
&&
\langle{0}| \epsilon^{ijk} u^i(0) C d^j(0)
[iD_\lambda \gamma_5 s_\gamma]^k(0)|{\Lambda,P}\rangle\nonumber\\
&& = P_\lambda\langle{0}|\epsilon^{ijk}u^i(0)cd^j(0)\gamma_5s_\gamma^k(0)|{\Lambda,P}\rangle-\langle{0}|\epsilon^{ijk}u^i(0)\stackrel{\leftrightarrow}{i D^\lambda}cd^j(0)\gamma_5s_\gamma^k(0)|{\Lambda,P}\rangle \,,
\nonumber \\ &&
\langle{0}| \epsilon^{ijk} u^i(0) C i \sigma_{\lambda\mu} \stackrel{\leftrightarrow}{i D^\mu} d^j(0)
s_\gamma^k(0)|{\Lambda,P}\rangle \nonumber \\ &&
= -P^\mu\langle{0}| \epsilon^{ijk} u^i(0) C i \sigma_{\lambda\mu}d^j(0)s^k(0) |{\Lambda,P}\rangle
+\langle{0}| \epsilon^{ijk} [u(0) C i \sigma_{\lambda\mu} d(0)]^{ij} iD^\mu s_\gamma^k(0) |{\Lambda,P}\rangle \,,
\nonumber \\ &&
\langle{0}| \epsilon^{ijk} u^i(0) C i \gamma_5\sigma_{\lambda\mu} \stackrel{\leftrightarrow}{i D^\mu} d^j(0)
s_\gamma^k(0)|{\Lambda,P}\rangle \nonumber \\ &&
= -P^\mu\langle{0}| \epsilon^{ijk} u^i(0) C i\gamma_5 \sigma_{\lambda\mu}d^j(0)s^k(0) |{\Lambda,P}\rangle
+\langle{0}| \epsilon^{ijk} [u(0) C i \gamma_5\sigma_{\lambda\mu} d(0)]^{ij} iD^\mu s_\gamma^k(0) |{\Lambda,P}\rangle\,,
\nonumber \\
&&
\langle{0}| \epsilon^{ijk} [u(0) C \gamma^\rho \stackrel{\leftrightarrow}{D}_\rho d(0) ]^{ij}s_\gamma^k(0)|{\Lambda,P}\rangle = 0
\,,
\nonumber \\
&&\langle{0}| \epsilon^{ijk} [u(0) C \{\gamma_\lambda i\stackrel{\leftrightarrow}{D}_\rho -
\gamma_\rho i\stackrel{\leftrightarrow}{D}_\lambda\} d(0)]^{ij}
s_\gamma^k(0)| {\Lambda,P}\rangle \nonumber
\\ &&
=- i \epsilon_{\lambda\rho\alpha\delta} [P^\alpha
\langle{0}| \epsilon^{ijk} u^i(0) C \gamma^\delta\gamma_5 d^j(0) s_\gamma^k(0)| {\Lambda,P}\rangle
-\langle{0}| \epsilon^{ijk} u^i(0) C \gamma^\delta\gamma_5 d^j(0)
[i D^\alpha s_\gamma]^k(0)| {\Lambda,P}\rangle] \,.
\nonumber \\
\end{eqnarray}
A simple but tedious calculation leads to the following equations:
\begin{eqnarray}
&\mathcal{A}_1^s+4\mathcal{A}_2^0+2\mathcal{A}_3^s=0,\, &-3\mathcal{A}_5^0=-\mathcal{A}_3^s+\mathcal{A}_4^0,\nonumber\\
&\mathcal{A}_1^0+\mathcal{A}_3^0=\mathcal{A}_1^s+\mathcal{A}_3^s+4\mathcal{A}_4^0+\mathcal{A}_2^0,\, &\mathcal{P}_1^s=\mathcal{P}_1^0,\nonumber\\
&\mathcal{P}_2^0=\mathcal{S}_2^0=0,\, &\mathcal{S}_1^s=\mathcal{S}_1^0,\nonumber\\
&-\mathcal{V}_1^0+\mathcal{V}_2^0+\mathcal{V}_3^0+4\mathcal{V}_4^0=0,\, &\mathcal{T}_1^0-3\mathcal{T}_2^0-2\mathcal{T}_3^0-\mathcal{T}_4^0=0\nonumber\\
&\mathcal{T}_1^0-2\mathcal{T}_3^0-\mathcal{T}_4^0+3\mathcal{T}_5^0
+6\mathcal{T}_7^0=0,\, &
\mathcal{T}_3^s-\mathcal{T}_4^0-3\mathcal{T}_7^0=0,\nonumber\\
&\mathcal{T}_3^0-\mathcal{T}_4^0=0\, &
\mathcal{V}_2^0-\mathcal{V}_3^0=\mathcal{A}_3^0-\mathcal{A}_3^s+\mathcal{A}_2^0,\nonumber\\
&2\mathcal{V}_5^0=-\mathcal{A}_3^0-\mathcal{A}_3^s+\mathcal{A}_2^0-2\mathcal{A}_5^0.&\label{eom}
\end{eqnarray}

Choosing $\mathcal{A}_1^0,\mathcal{A}_3^0,\mathcal{A}_1^s,\mathcal{A}_3^s,\mathcal{V}_1^0,\mathcal{V}_2^0,\mathcal{T}_1^0,\mathcal{T}_2^0,\mathcal{P}_1^0,
\mathcal{S}_1^0$ as the independent parameters, the other ones can be expressed with them:
\begin{eqnarray}
&\mathcal{A}_2^0=\frac14(-\mathcal{A}_1^s-2\mathcal{A}_3^s),&\mathcal{A}_4^0=\frac{1}{16}(4\mathcal{A}_1^0+4\mathcal{A}_3^0-3\mathcal{A}_1^s-2\mathcal{A}_3^s),\nonumber\\
&\mathcal{A}_5^0=\frac{1}{48}(-4\mathcal{A}_1^0+4\mathcal{A}_3^0+3\mathcal{A}_1^s+\mathcal{A}_3^s), &\mathcal{V}_3^0=\frac{1}{4}(-\mathcal{A}_3^0+\mathcal{A}_1^s+2\mathcal{A}_3^s+\mathcal{V}_2^0,\nonumber\\
&\mathcal{V}_4^0=\frac{1}{16}(4\mathcal{A}_3^0-\mathcal{A}_1^s-6\mathcal{A}_3^s+4\mathcal{V}_1^0-8\mathcal{V}_2^0),
&\mathcal{V}_5^0=\frac{1}{48}(4\mathcal{A}_1^0-36\mathcal{A}_3^0-9\mathcal{A}_1^s-38\mathcal{A}_3^s),\nonumber\\
&\mathcal{T}_3^0=\frac{1}{3}(\mathcal{T}_1^0-3\mathcal{T}_2^0),&\mathcal{T}_4^0
=\frac{1}{3}(\mathcal{T}_1^0-3\mathcal{T}_2^0),\nonumber\\
&\mathcal{T}_5^0=-6\mathcal{T}_2^0,&\mathcal{T}_7^0=0,\nonumber\\
&\mathcal{P}_1^s=\mathcal{P}_1^0,&\mathcal{P}_2^0=0\nonumber\\
&\mathcal{S}_1^s=\mathcal{S}_1^0,&\mathcal{S}_2^0=0.
\end{eqnarray}

There are so far $10$ independent parameters to be determined to give the expressions of the LCDAs of $\Lambda$. To this end, the six coupling constants defined by the following matrix elements of a three-quark operator with a covariant derivative are introduced
\begin{eqnarray}
&& \langle{0}| \epsilon^{ijk} \left[u^i(0) C \gamma_5\!\not\!{z} d^j(0)\right]\!\not\!{z} (iz\vec{D}s^k)(0)| {\Lambda(P)}\rangle = f_\Lambda
A_1^s(P\cdot z)^2 \!\not\!{z} \Lambda(P)_\gamma\,, \nonumber \\
&& \langle{0}| \epsilon^{ijk} \left[u^i(0) C \!\not\!{z} iz\stackrel{\leftrightarrow}{D}d^j(0)\right] \gamma_5\!\not\!{z} s^k(0)| {\Lambda(P)}\rangle =-f_\Lambda
A_1^q(P\cdot z)^2 \!\not\!{z} \Lambda(P)_\gamma\,, \nonumber \\
&& \langle{0}| \epsilon^{ijk} \left[u^i(0) C \gamma_5\gamma^u d^j(0)\right] \!\not\!{z}\gamma^u (iz\vec{D}s^k)(0)| {\Lambda}\rangle = \lambda_1f_1^s
(P\cdot z)M \!\not\!{z} \Lambda(P)_\gamma\,, \nonumber \\
&& \langle{0}| \epsilon^{ijk} \left[u^i(0) C \gamma^uiz\stackrel{\leftrightarrow}{D} d^j(0)\right] \, \gamma_5 \!\not\!{z}\gamma_\mu s^k(0)| {\Lambda}\rangle =-\lambda_1
f_1^q(P\cdot z)(P\cdot z) \!\not\!{z} \Lambda(P)_\gamma\,, \nonumber \\
&& \langle{0}| \epsilon^{ijk} \left[u^i(0) C\sigma_{\mu\nu} iz\stackrel{\leftrightarrow}{D}d^j(0)\right] \gamma_5\!\not\!{z}\sigma^{\mu\nu} s^k(0)| {\Lambda(P)}\rangle =-\lambda_3
f_3^q(P\cdot z)M\!\not\!{z} \Lambda(P)_\gamma\,, \nonumber \\
&& \langle{0}| \epsilon^{ijk} \left[u^i(0)iP^\nu C\sigma^{\mu\nu} iz\stackrel{\leftrightarrow}{D}d^j(0)\right] \, \gamma_5 \!\not\!{z} \gamma_\mu s^k)(0)| {\Lambda(P)}\rangle =-\lambda_3
f_4^q(P\cdot z)M^2 \!\not\!{z} \Lambda(P)_\gamma\,.\nonumber\\
\label{def-nonlocal2}
\end{eqnarray}

Another four coupling constants are defined by the leading order local operator matrix elements which have been calculated in Ref. \cite{DAs-sig}
\begin{eqnarray}
&& \langle{0}| \epsilon^{ijk} \left[u^i(0) C\gamma_5 \!\not\!{z}
d^j(0)\right] \!\not\!{z} s^k(0)| {P}\rangle =   f_{\Lambda}
(pz) \!\not\!{z} \Lambda(P)\,,  \nonumber \\
&&\langle{0}| \epsilon^{ijk} \left[u^i(0) C\gamma_5\gamma_\mu
d^j(0)\right]\, \gamma^\mu s^k(0)| {P}\rangle = \lambda_1 M
\Lambda(P) \,,
\nonumber \\
&&\langle{0}| \epsilon^{ijk} \left[u^i(0) C\gamma_5 d^j(0)\right] \,
s^k(0)|{P}\rangle = \lambda_2 M \Lambda(P) \,,
\nonumber\\
&&\langle{0}| \epsilon^{ijk} \left[u^i(0) Cd^j(0)\right]\,\gamma_5
s^k(0)|{P}\rangle=\lambda_3 M^2\Lambda(P) \,.\label{def-local}
\end{eqnarray}
The relations between the local nonperturbative parameters $\mathcal V_i^{0}, \mathcal A_i^{0,s}$, $\mathcal T_i^{0},\mathcal P_i^{0},\mathcal S_i^{0}$ and the coupling constants defined in Eqs. (\ref{def-nonlocal2}) and (\ref{def-local}) are give as:
\begin{eqnarray}
&f_{\Lambda}=\mathcal A_1^0,&\lambda_1=\mathcal A_1^0+4\mathcal A_3^0,\nonumber\\
&\lambda_2=\mathcal P_1^0,&\lambda_3=\mathcal S_1^0,\nonumber\\
&f_\Lambda A_1^s=\mathcal A_1^s,&f_\Lambda A_1^q=-\mathcal V_1^0,\nonumber\\
&\lambda_1 f_1^s=\mathcal A_1^s+4\mathcal A_3^s,&\lambda_1f_1^q=\mathcal V_1^0-2\mathcal V_2^0-4 \mathcal V_3^0,\nonumber\\
&\lambda_3 f_3^q=6\mathcal T_1^0-2\mathcal T_2^0-24\mathcal T_3^0-10\mathcal T_4^0,&\lambda_3f_4^q=3\mathcal T_1^0+\mathcal T_2^0+6\mathcal T_3^0-\mathcal T_4^0+12\mathcal T_7^0.
\end{eqnarray}
With the above preparations we can express all the independent parameters by the nonperturbative coupling constants defined in Eqs. (\ref{def-nonlocal2}) and (\ref{def-local}):
\begin{eqnarray}
\mathcal A_1^0&=&f_\Lambda, \hspace{1cm}\mathcal A_3^0=-\frac{1}{4}(f_\Lambda-\lambda_1),\hspace{1cm}\mathcal A_1^s=f_\Lambda A_1^s,\nonumber\\
\mathcal A_3^s&=&-\frac{1}{4}(A_1^sf_\Lambda-f_1^s\lambda_1),\hspace{3cm} \mathcal V_1^0=-f_\Lambda A_1^q,\nonumber\\
\mathcal V_2^0&=&\frac{1}{12}(-2f_\Lambda+A_1^sf_\Lambda+2\lambda_1-2f_1^q\lambda_1-3f_1^s\lambda_1-2f_\Lambda A_1^q),\nonumber\\
\mathcal T_1^0&=&\frac{1}{32}(f_3^q+8f_4^q), \hspace{3cm}\mathcal T_2^0=\frac{1}{192}(7f_3^q+8f_4^q),\nonumber\\
\mathcal P_1^0&=&\lambda_2,\hspace{1cm}\mathcal S_1^0=\lambda_3.
\end{eqnarray}

A tedious calculation shows that coefficients in Eqs. (\ref{contwist3})-(\ref{contwist6}) can be expressed to the next-to-leading order conformal spin accuracy as
\begin{eqnarray}
\phi_3^0&=&\phi_6^0={\mathcal A}_1^0,\hspace{4cm}\psi_4^0=\psi_5^0=-2{\mathcal A}_3^0,\nonumber\\
\phi_4^0&=&\phi_5^0=-{\mathcal A}_1^0-2{\mathcal A}_3^0,\hspace{2.5cm}{\xi}_4^0=\frac12({\mathcal P}_1^0+2{\mathcal S}_1^0),\nonumber\\
{\xi'}_4^0&=&-\frac12({\mathcal P}_1^0-2{\mathcal S}_1^0),\hspace{2.9cm}{\xi}_5^0={\mathcal P}_1^0+{\mathcal S}_1^0,\nonumber\\
{\xi'}_5^0&=&{\mathcal P}_1^0-{\mathcal S}_1^0,\hspace{3.8cm}t_1^0=t_2^0=t_5^0=t_6^0=0
\end{eqnarray}
for leading order and
\begin{eqnarray}
\phi_3^+&=&\frac{7}{2}{\mathcal A}_1^0-\frac{21}{2}{\mathcal A}_1^s,\hspace{2cm} \phi_6^+=-2{\mathcal A}_1^0+12({\mathcal A}_1^s+{\mathcal A}_2^0+{\mathcal A}_4^0+{\mathcal A}_5^0),\nonumber\\
\phi_3^-&=&-\frac{21}{2}{\mathcal V}_1^0,\hspace{3cm} \phi_4^-=\frac{15}{2}({\mathcal V}_1^0+2{\mathcal V}_3^0),\nonumber\\
\phi_5^-&=&-10({\mathcal V}_1^0-2{\mathcal V}_3^0+2{\mathcal V}_4^0-2{\mathcal V}_5^0),\hspace{2cm} \phi_6^-=-6({\mathcal V}_1^0-2{\mathcal V}_2^0+2{\mathcal V}_4^0+2{\mathcal V}_5^0),\nonumber\\
\phi_4^+&=&-\frac{3}{2}({\mathcal A}_1^0-5{\mathcal A}_1^s-10{\mathcal A}_2^0+6{\mathcal A}_3^0-10{\mathcal A}_3^s),\nonumber\\
\phi_5^+&=&-5{\mathcal A}_1^0-10{\mathcal A}_3^0+20{\mathcal A}_3^s+20{\mathcal A}_4^0-20{\mathcal A}_5^0,\nonumber\\
\xi_4^-&=&\frac{15}{2}{\mathcal S}_1^0-\frac{45}{2}{\mathcal S}_1^s-15{\mathcal T}_1^0+15{\mathcal T}_2^0+30{\mathcal T}_3^0,\nonumber\\
{\xi'}_4^-&=&-\frac{15}{4}{\mathcal P}_1^0+\frac{45}{2}{\mathcal P}_1^s+\frac{15}{2}{\mathcal S}_1^0-\frac{45}{2}{\mathcal S}_1^s-15{\mathcal T}_1^0+15{\mathcal T}_2^0+60{\mathcal T}_3^0,\nonumber\\
\xi_4^+&=&-3{\mathcal P}_1^0+15{\mathcal P}_1^s+\frac{3}{2}{\mathcal S}_1^0-\frac{15}{2}{\mathcal S}_1^s-15{\mathcal T}_1^0+15{\mathcal T}_2^0+30{\mathcal T}_3^0,\nonumber\\
{\xi'}_4^+&=&-\frac{3}{4}{\mathcal P}_1^0+\frac{15}{2}{\mathcal P}_1^s+\frac{3}{2}{\mathcal S}_1^0-\frac{15}{2}{\mathcal S}_1^s-15{\mathcal T}_1^0+15{\mathcal T}_2^0+60{\mathcal T}_3^0,\nonumber\\
\xi_5^-&=&-25{\mathcal P}_1^0+40{\mathcal P}_2^0+20{\mathcal P}_1^s+45{\mathcal S}_1^0-20{\mathcal S}_1^s+\frac{315}{4}{\mathcal T}_1^0+20{\mathcal T}_3^0
-15{\mathcal T}_5^0-40{\mathcal T}_7^0,\nonumber\\
{\xi'}_5^-&=&-5{\mathcal P}_1^0+40{\mathcal P}_2^0+20({\mathcal P}_1^s+45{\mathcal S}_1^0)-20{\mathcal S}_1^s,\nonumber\\
\xi_5^+&=&-10{\mathcal P}_1^0+40{\mathcal P}_2^0+20{\mathcal P}_1^s+\frac{105}{4}{\mathcal T}_1^0-5{\mathcal T}_5^0,\nonumber\\
{\xi'}_5^+&=&-20({\mathcal P}_1^0+40{\mathcal P}_2^0)+20({\mathcal P}_1^s+\frac{105}{2}{\mathcal T}_1^0+20{\mathcal T}_3^0-10{\mathcal T}_5^0-40{\mathcal T}_7^0),\nonumber\\
\psi_4^+&=&-\frac{15}{2}({\mathcal A}_3^s+{\mathcal V}_3^0)+\frac{9}{2}{\mathcal A}_3^0,\hspace{3.5cm}\psi_4^-=-\frac{15}{2}({\mathcal A}_3^0+{\mathcal V}_3^0-3{\mathcal A}_3^s),\nonumber\\
\psi_5^+&=&-10{\mathcal A}_3^s+20{\mathcal A}_5^0-10{\mathcal V}_3^0-20{\mathcal V}_5^0,\hspace{2cm}\psi_5^-=10{\mathcal A}_3^0+30{\mathcal A}_3^s-60{\mathcal A}_5^0-10{\mathcal V}_3^0-20{\mathcal V}_5^0,\nonumber\\
t_1^-&=&-\frac{21}{2}{\mathcal T}_1^0,\hspace{6.0cm} t_2^-=\frac{15}{2}{\mathcal T}_1^0+30{\mathcal T}_3^0-15{\mathcal T}_4^0,\nonumber\\
t_6^-&=&-12({\mathcal T}_1^0+{\mathcal T}_4^0+{\mathcal T}_5^0),\nonumber\\
t_5^-&=&-6{\mathcal P}_1^0+16{\mathcal P}_2^0+8{\mathcal P}_1^s+4{\mathcal S}_1^0-2{\mathcal S}_1^s+\frac{63}{4}{\mathcal T}_1^0+2{\mathcal T}_2^0+2{\mathcal T}_3^0-3{\mathcal T}_5^0-4{\mathcal T}_7^0,\nonumber\\
t_1^+&=&t_2^+=t_5^+=t_6^+=0.
\nonumber\\
\end{eqnarray}
for the next-to-leading order.

\section{numerical analysis of the sum rules for the nonperturbative parameters}\label{sec:sumrule}

The next step is to determine the nonperturbative parameters defined in Eq. (\ref{def-nonlocal2}) which give the estimations of the conformal expansion coefficients of the light-cone distribution amplitudes with the chiral field representation. We use two-point QCD sum rules \cite{svzsum} to reach this aim in this section. This method has been used to calculate the higher moments of the baryon LCDAs thirty years ago \cite{Cheryak-sum}. It starts from the two-point correlation function. On one side the phenomenon representation is obtained by inserting a complete set of hadron states that have the same quantum numbers as the $\Lambda$ baryon. On the other side, the correlation function can be calculated directly at quark level in aid of the operator product expansion (OPE) technique. The nonperturbative effects are included into the so-called vacuum condensates. Then matching the two sides with the help of quark-hadron duality, the nonperturbative parameters can be expressed by the integral of the spectral density plus the vacuum condensates. In the calculations we consider vacuum condensates up to dimension $6$. The detailed processes are referred to the previous work \cite{DAs-sighigherorder}. In compliance with the standard procedure of the QCD sum rule method, we arrive at the following results:

\begin{itemize}
\item
The sum rule for $A_1^s$ is
\begin{eqnarray}
2f_\Lambda^2A_1^se^{-M^2/M_B^2}=\int_{m_s^2}^{s_0}e^{-s/M_B^2}\rho(s)ds+\Pi^{cond.},\label{sumruleofv1s}
\end{eqnarray}
with
\begin{eqnarray}
\rho(s)&=&\frac{1}{5\times3\times2^6\pi^4}s(1-x)^5(1+2x)+\frac{\langle g^2G^2\rangle}{3\times2^7\pi^4}\frac{1}{s}x(1-x)(1-5x+x^2),
\end{eqnarray}
and
\begin{equation}
\Pi^{cond.}=\frac{m(m_0^2-2m_s^2)}{3^2\times2^4\pi^2}\langle \bar ss\rangle\frac{1}{M_B^2}-\frac{m_s}{3^2\times2^5\pi^2}\langle \bar sg\cdot \sigma Gs\rangle\frac{1}{M_B^2}(1+\frac{m_s^2}{M_B^2}),
\end{equation}
where $x=m_s^2/s$, $m_s$ is the strange quark mass, $M$ is the mass of $\Lambda$ and $M_B^2$ is the Borel parameter.
\item
The sum rule for $A_1^q$ is
\begin{eqnarray}
2f_\Lambda^2A_1^qe^{-M^2/M_B^2}=\int_{m_s^2}^{s_0}e^{-s/M_B^2}\rho(s)ds+\Pi^{cond.},
\end{eqnarray}
where
\begin{equation}
\rho(s)=\frac{\langle g^2G^2\rangle}{3^2\times2^7\pi^4}\frac{1}{s}x(1-x)^3,
\end{equation}
and
\begin{equation}
\Pi^{cond.}=-\frac{\langle \bar s\sigma\cdot Gs\rangle}{3^2\times2^5\pi^2}\frac{m_s}{M_B^2}(1+\frac{m_s^2}{M_B^2}).
\end{equation}
\item
The sum rule for $f_1^s$ is
\begin{eqnarray}
\lambda_1^2M^2f_1^{s}e^{-M^2/M_B^2}=\int_{m_s^2}^{s_0}e^{-s/M_B^2}\rho(s)ds+\Pi^{cond.},
\end{eqnarray}
where
\begin{eqnarray}
\rho^{}(s)&=&-\frac{s^2}{5\times3\times2^8\pi^4}\{5(1-x)(1+x)(1-8x+x^2)-14(1-x)^5\nonumber\\
&&+60x^2\ln x\}+\frac{\langle g^2G^2\rangle}{3^2\times2^9\pi^4}(1-x)^2(9+2x),
\end{eqnarray}
\begin{eqnarray}
\Pi^{cond.}&=&-\frac16(2-2\frac{m_0^2}{M_B^2}-\frac{m_0^2m_s^2}{M_B^4})\langle \bar qq\rangle^2e^{-\frac{m_s^2}{M_B^2}}-\frac{m_s}{48\pi^2}\langle\bar ss\rangle (M_B^2-3m_0^2+6m_s^2)\nonumber\\
&&+\frac{m_s\langle\bar sg\sigma\cdot Gs\rangle}{3^2\times2^6\pi^2}(3-2\frac{m_s^2}{M_B^2})\frac{1}{M_B^2}.
\end{eqnarray}
\item
The sum rule for $f_1^q$ is
\begin{eqnarray}
-\lambda_1^2M^2f_1^{q}e^{-M^2/M_B^2}=\int_{m_s^2}^{s_0}e^{-s/M_B^2}\rho(s)ds+\Pi^{cond.},
\end{eqnarray}
where
\begin{eqnarray}
\rho(s)&=&\frac{s^2}{5\times3\times2^9\pi^4}\{[(1-x)(3-27x-47x^2+13x^3-2x^4)\nonumber\\
&&-60x^2\ln x]\}+\frac{\langle g^2G^2\rangle}{3^3\times2^9\pi^4}(1-x)^2(19-28x),
\end{eqnarray}
\begin{eqnarray}
\Pi^{cond.}&=&-\frac{m_s}{3\times2^4\pi^2}\langle\bar ss\rangle(2M_B^2-m_0^2+2m_s^2).
\end{eqnarray}
\item
The sum rule for $f_3^q$ is
\begin{eqnarray}
-\lambda_3^2M^2f_3^{q}e^{-M^2/M_B^2}=\int_{m_s^2}^{s_0}e^{-s/M_B^2}\rho(s)ds+\Pi^{cond.},
\end{eqnarray}
where
\begin{eqnarray}
\rho(s)&=&-\frac{s^2}{5\times3\times 2^8\pi^4}\{[(1-x)(3-27x-47x^2+13x^3-2x^4)\nonumber\\
&&-60x\ln x]\}-\frac{\langle g^2G^2\rangle}{3^2\times2^9\pi^4}(1-x)^2(5-8x),
\end{eqnarray}
\begin{eqnarray}
\Pi^{cond.}&=&-\frac{m_s}{3\times2^5\pi^2}\langle\bar ss\rangle(2M_B^2+m_0^2-2m_s^2)-\frac{5m_s\langle \bar sq\sigma\cdot Gs\rangle}{3\times2^6\pi^2}(1+\frac{m_s^2}{M_B^2}).
\end{eqnarray}
\item
The sum rule for $f_4^q$ is
\begin{eqnarray}
-\lambda_3^2M^3f_4^{q}e^{-M^2/M_B^2}=\int_{m_s^2}^{s_0}e^{-s/M_B^2}\rho(s)ds+\Pi^{cond.},
\end{eqnarray}
where
\begin{eqnarray}
\rho(s)&=&-\frac{m_s}{3\times2^9\pi^4}s^2\{(1-x)(3+47x+11x^2-x^3)+12x(2+3x)\ln x]\}\nonumber\\
&&-\frac{m_s\langle g^2G^2\rangle}{3^3\times2^{11}\pi^4}\frac{1}{x}[(1-x)(48+97x-11x^2+4x^3)+138x\ln x]\nonumber\\
&&+\frac{m_s\langle g^2G^2\rangle}{2^9\pi^4}(1-x)^2,
\end{eqnarray}
\begin{eqnarray}
\Pi^{cond.}&=&\frac{\langle\bar ss\rangle}{3^2\times2^5\pi^2}(2M_B^2-m_0^2+2m_s^2)M_B^2-\frac{\langle \bar sg\sigma\cdot Gs\rangle }{3^2\times2^8\pi^2}(M_B^2-m_s^2).
\end{eqnarray}

\end{itemize}

Before getting the sum rules above, we perform the Borel transformation on the squared transfer momentum to make the sum rules more reliable. Therefore we first need to determine the working window of the Borel parameter, which is obtained by requiring that both the higher order resonance contributions are subdominant in comparison with the pole contribution and the higher dimension contributions have a good convergence. In detail, we choose the lower limit of the Borel mass by requiring that the condensate contributions $\Pi^{cond.}$ are less than $30\%$ and have good convergence with the increment of dimension. At the same time, the resonance contributions are less than that of the pole terms which give the upper limit of $M_B^2$. To satisfy the above criterion we set the Borel mass for different sum rules which are presented in Tab.\ref{tab:sumrule}.

Another important parameter in the QCD sum rules is the threshold $s_0$, by choosing which the higher resonance and continuum contributions can be represented by the integral of the spectral density with the help of quark-hadron duality. The threshold is usually connected with the first excited state which has the same quantum number as the concerned composite particle. It is also required that the sum rule does not dependent on the threshold very much. In compliance with the above requirements, in the numerical analysis we use $2.5\;\mbox{GeV}^2\leq s_0\leq 2.7\;\mbox{GeV}^2$.

\begin{figure}
\begin{minipage}{8cm}
\epsfxsize=7cm \centerline{\epsffile{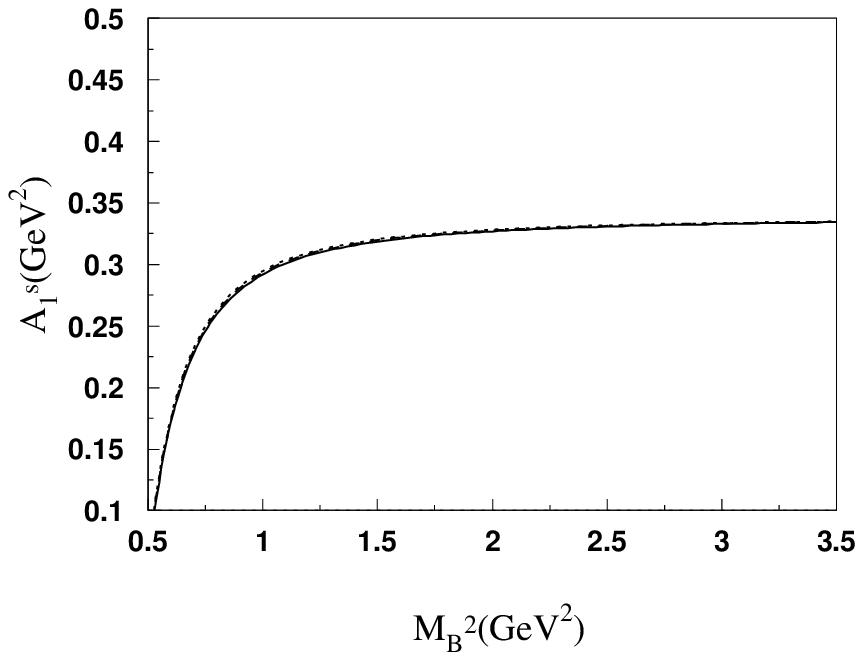}}
\end{minipage}
\begin{minipage}{8cm}
\epsfxsize=7cm \centerline{\epsffile{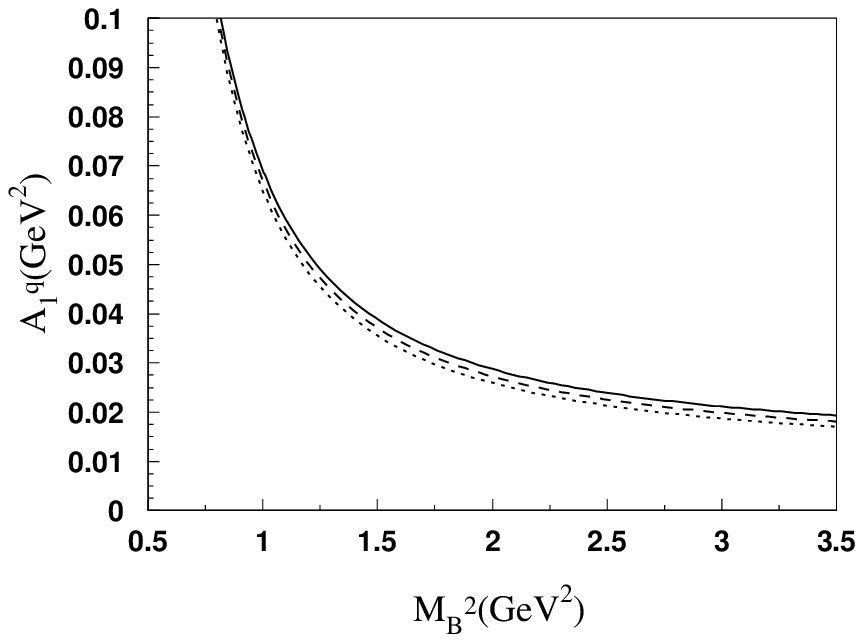}}
\end{minipage}
\begin{minipage}{8cm}
\epsfxsize=7cm \centerline{\epsffile{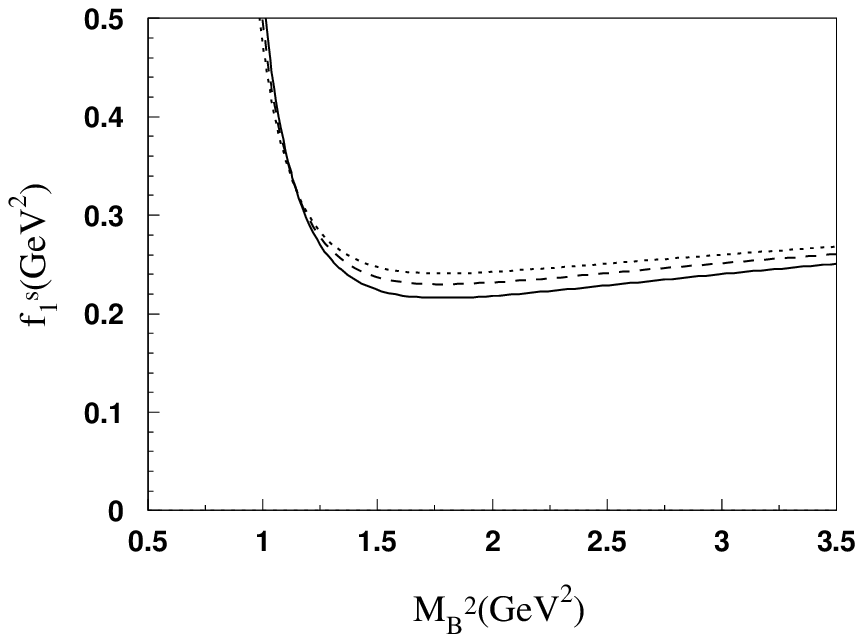}}
\end{minipage}
\begin{minipage}{8cm}
\epsfxsize=7cm \centerline{\epsffile{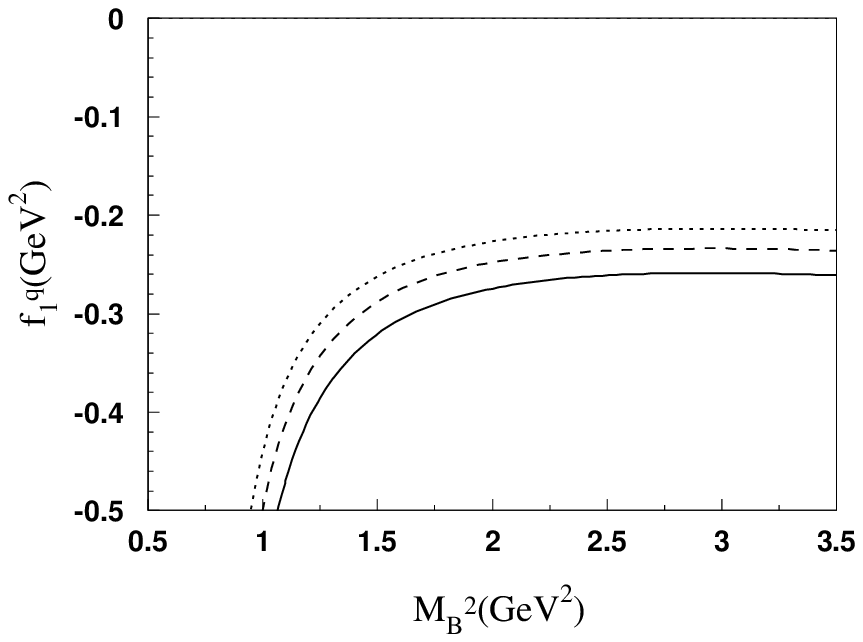}}
\end{minipage}
\begin{minipage}{8cm}
\epsfxsize=7cm \centerline{\epsffile{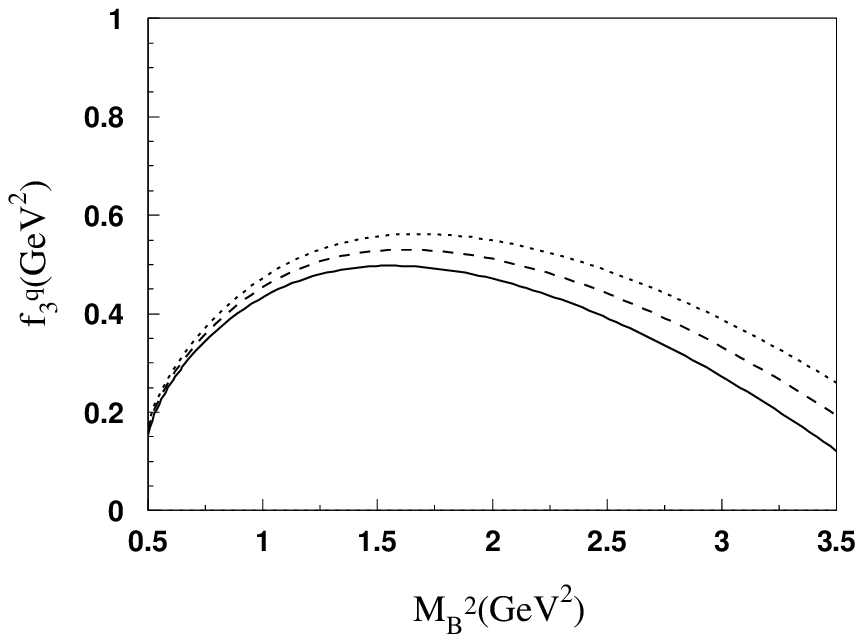}}
\end{minipage}
\begin{minipage}{8cm}
\epsfxsize=7cm \centerline{\epsffile{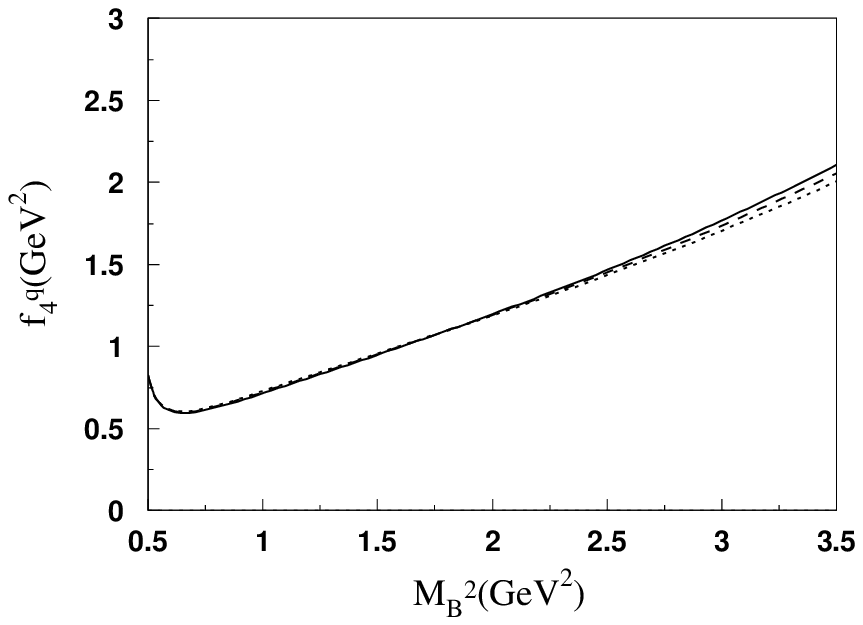}}
\end{minipage}
\caption{\quad Borel working window of the nonperturbative parameters with threshold $s_0$. The solid, the dashed, and the dotted lines correspond to $2.6\;\mbox{GeV}^2, 2.5\;\mbox{GeV}^2, 2.7\;\mbox{GeV}^2$, respectively.} \label{fig:1}
\end{figure}

Finally, the inputs of the vacuum condensates we used are the standard values: $a=-(2\pi)^2\langle\bar
uu\rangle=0.55\; \mbox{GeV}^{3}$, $b=(2\pi)^2\langle\alpha_sG^2/\pi\rangle=0.47\; \mbox{GeV}^{4}$, $a_s=-(2\pi)^2\langle\bar ss\rangle=0.8a$, $\langle\bar
ug_c\sigma\cdot Gu\rangle=m_0^2\langle\bar uu\rangle$, and $m_0^2=0.8\; \mbox{GeV}^{2}$. The mass of the strange quark is used as $m_s=0.15\,\mbox{GeV}$.
The baryon mass is adopted the central value of $\Lambda$ presented by the Particle Data Group (PDG) \cite{PDG}: $M_{\Lambda}=1.116\mbox{GeV}$. The sum rules dependent on the Borel parameters are shown in Fig.\ref{fig:1}. The estimations of the nonperturbative parameters are shown in Tab.\ref{tab:sumrule}. In the numerical analysis the errors of the coupling constants come from both the variation of the Borel mass and the threshold $s_0$. It is noticed that the figures show that not all the coupling constants increase with the increment of $s_0$. This lies in the fact that our definitions in Eq.(\ref{def-nonlocal2}) are related to the couplings of leading order ones. That is to say, the results are the obtained expressions divided by the squared coupling constants of the leading order.

\begin{table}
\renewcommand{\arraystretch}{1.1}
\caption{Results from QCD sum rules of the nonperturbative parameters with $2.5\mbox{GeV}^2\leq s_0\leq 2.7\mbox{GeV}^2$.}
\begin{center}
\begin{tabular}{|l|l|l|l|l|l|l|}
\hline
Parameter & $A_1^s$ & $A_1^q$ & $f_1^s$ & $f_1^q$ &$f_3^q$ &$f_4^q$\\
\hline
$M_B^2(GeV^2)$ & $1.0\sim1.5$ & $1.5\sim2$ &$0.8\sim1.0$ & $1.5\sim2.0$ &$0.8\sim1.1$& $1.5\sim2.0$\\
\hline
Results & $0.31\pm0.01$ & $0.032\pm0.006$ &$0.23\pm0.01$ & $-0.23\pm0.03$ & $0.43\pm0.07$& $1.07\pm0.12$\\
\hline
\end{tabular}
\end{center} \label{tab:sumrule}
\end{table}

\section{Explicit expressions of the $\Lambda$ LCDAs}\label{sec:result}

\begin{figure}
\begin{minipage}{8cm}
\epsfxsize=7cm \centerline{\epsffile{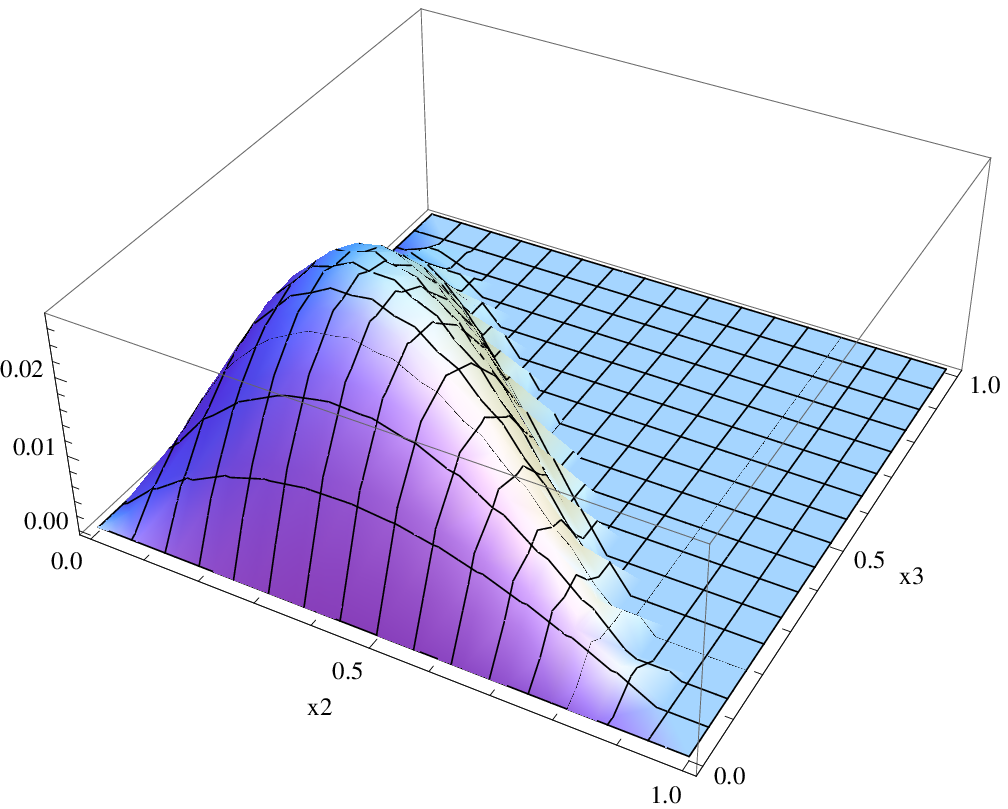}}
\end{minipage}
\begin{minipage}{8cm}
\epsfxsize=7cm \centerline{\epsffile{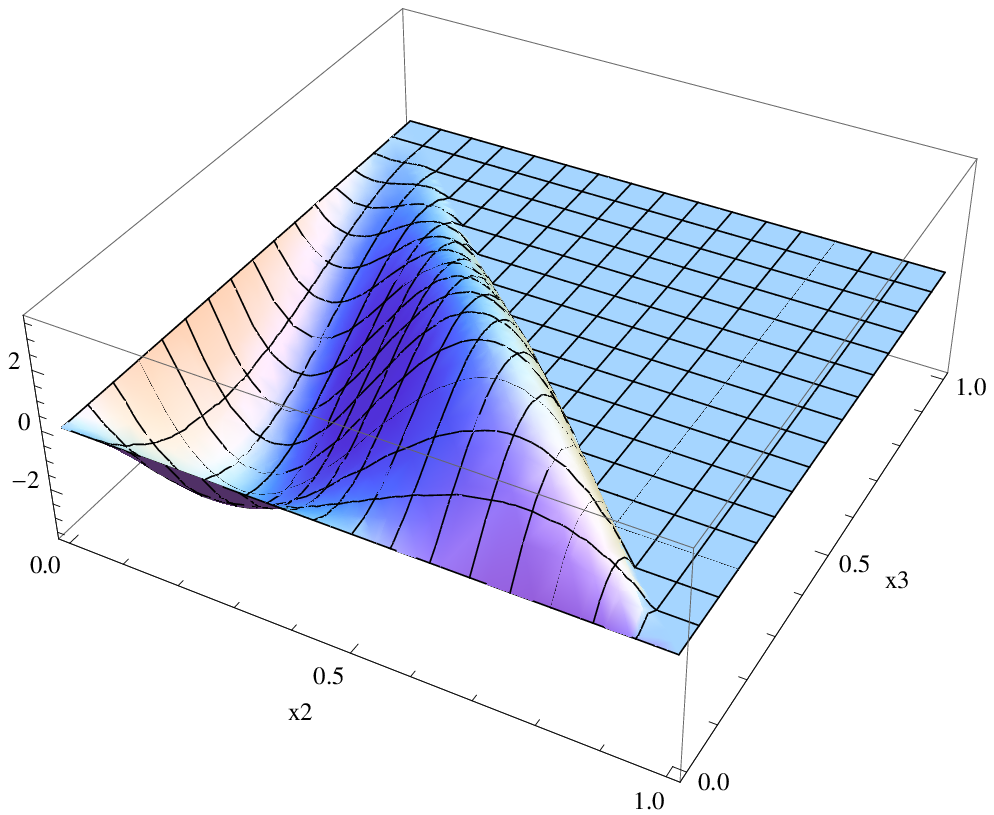}}
\end{minipage}
\begin{minipage}{8cm}
\epsfxsize=7cm \centerline{\epsffile{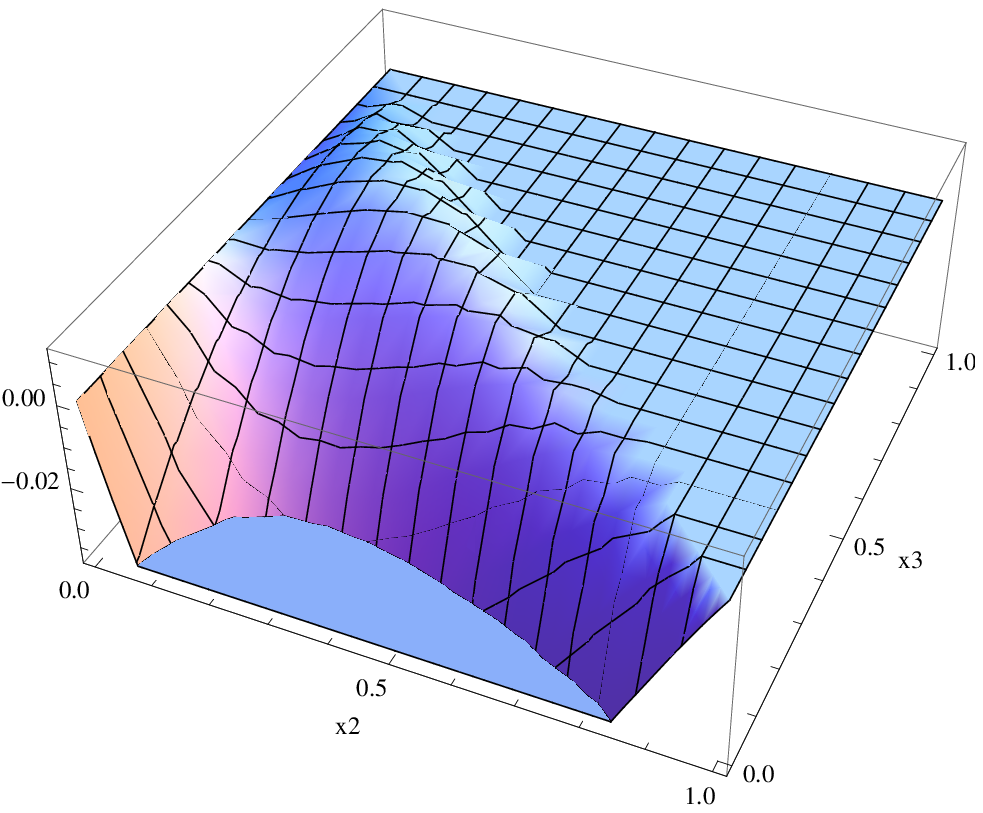}}
\end{minipage}
\begin{minipage}{8cm}
\epsfxsize=7cm \centerline{\epsffile{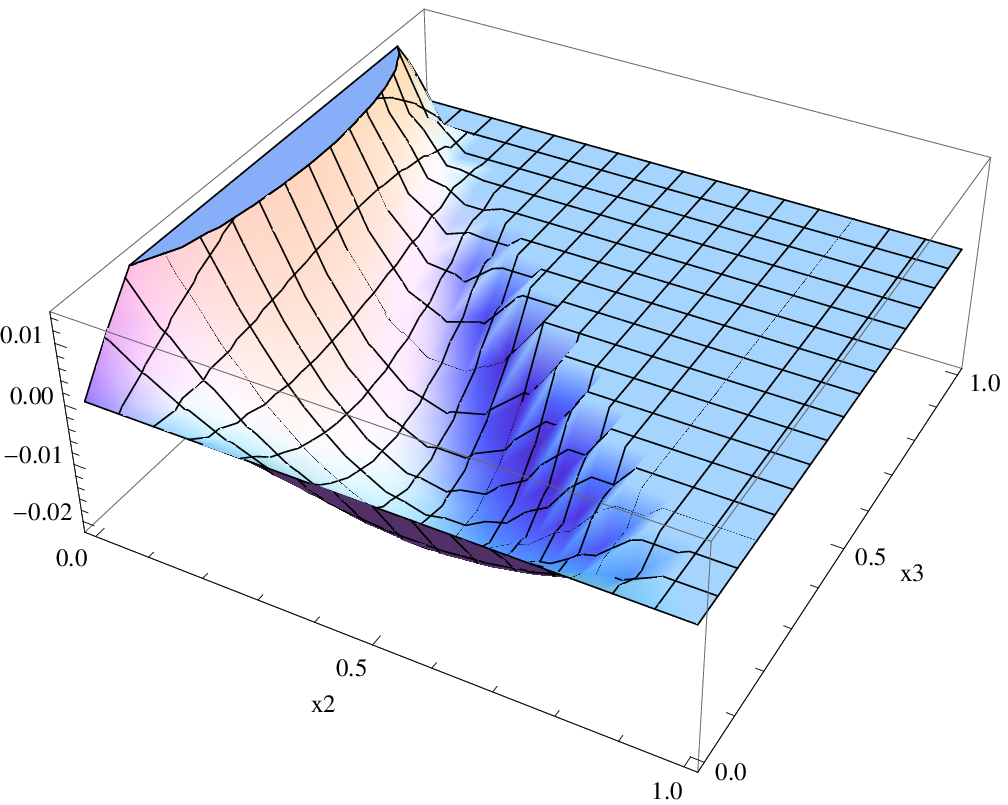}}
\end{minipage}
\caption{\quad Twist-$3$ distribution amplitudes $\Phi_3(x_i)$, $t_1(x_i)$ (up) and Twist-$4$ distribution amplitudes $\Phi_4(x_i)$, $\Psi_4(x_i)$ (down).} \label{phi-figure}
\end{figure}

Now we can write down the explicit expressions of the $\Lambda$ baryon LCDAs defined in Eq.(\ref{da-def}). By considering expressions in (\ref{contwist3}) to (\ref{contwist6}), we first plot the twist-$3$ distribution amplitudes $\Phi_3(x_i)$, $t_1(x_i)$ and two of the twist-$4$ distribution amplitudes $\Phi_4(x_i)$, $\Psi_4(x_i)$ in Fig. \ref{phi-figure} as an example.

For the definition in (\ref{da-deftwist}), our results are listed as follows: Twist-$3$ distribution amplitudes of $\Lambda$ are
\begin{eqnarray}
V_1(x_i)&=&120x_1x_2x_3(x_1-x_2)\phi_3^-,\hspace{1.0cm}A_1(x_i)=-120x_1x_2x_3[\phi_3^0+\phi_3^+(1-3x_3)],\nonumber\\
T_1(x_i)&=&120x_1x_2x_3[t_1^0+t_1^-(x_1-x_2)+t_1^+(1-3x_3)].
\end{eqnarray}
Twist-$4$ distribution amplitudes are
\begin{eqnarray}
S_1(x_i)&=&6(x_2+x_1)x_3(\xi_4^0+\xi_4^{'0})+6(x_2^2+x_1^2-(x_2+x_1)x_3)x_3(\xi_4^-+\xi_4^{'-})\nonumber\\
&&+6(x_2+x_1-10x_1x_2)x_3(\xi_4^++\xi_4^{'+}),\nonumber\\
P_1(x_i)&=&6(x_2+x_1)x_3(\xi_4^0-\xi_4^{'0})+6(x_2^2+x_1^2-(x_2+x_1)x_3)x_3(\xi_4^--\xi_4^{'-})\nonumber\\
&&+6(x_2+x_1-10x_1x_2)x_3(\xi_4^+-\xi_4^{'+}),\nonumber\\
V_2(x_i)&=&24x_1x_2(x_1-x_2)\phi_4^-,\hspace{2cm}A_2(x_i)=-24x_1x_2[\phi_4^0+\phi_4^+(1-5x_3)],\nonumber\\
V_3(x_i)&=&12x_3(x_1-x_2)\psi_4^0+12(x_1^2-x_2^2-(x_1-x_2)x_3)x_3\psi_4^-+12(x_1-x_2)x_3\psi_4^-,\nonumber\\
A_3(x_i)&=&-12x_3(1-x_3)\psi_4^0-12[(x_1^2+x_2^2)x_3-(x_1+x_2)x_3^2]\psi_4^-\nonumber\\
&&-12x_3(x_1+x_2-10x_1x_2)\psi_4^+,\nonumber\\
T_2(x_i)&=&24x_1x_2[t_2^0+t_2^-(x_1-x_2)+t_2^+(1-5x_3)],\nonumber\\
T_3(x_i)&=&6x_3(x_2-x_1)(\xi_4^0+\xi_4^{'0})+6(x_1^2-x_2^2-x_3(x_2-x_1))x_3(\xi_4^-+\xi_4^{'-})\nonumber\\
&&+6(x_2-x_1)x_3(\xi_4^++\xi_4^{'+}),\nonumber\\
T_7(x_i)&=&-6x_3(x_2-x_1)(\xi_4^0-\xi_4^{'0})-6(x_2^2-x_1^2-x_3(x_2-x_1))x_3(\xi_4^--\xi_4^{'-})\nonumber\\
&&-6(x_2-x_1)x_3(\xi_4^+-\xi_4^{'+}).
\end{eqnarray}
Twist-$5$ distribution amplitudes are
\begin{eqnarray}
S_2(x_i)&=&\frac32(x_1+x_2)(\xi_5^0+\xi_5^{'0})+\frac32(x_1+x_2-2(x_1^2+x_2^2))(\xi_5^++\xi_5^{'+})\nonumber\\
&&+\frac32(2x_1x_2-(x_1+x_2)x_3)(\xi_5^-+\xi_5^{'-}),\nonumber\\
P_2(x_i)&=&\frac32(x_1+x_2)(\xi_5^0-\xi_5^{'0})+\frac32(x_1+x_2-2(x_1^2+x_2^2))(\xi_5^+-\xi_5^{'+})\nonumber\\
&&+\frac32(2x_1x_2-(x_1+x_2)x_3)(\xi_5^--\xi_5^{'-}),\nonumber\\
V_4(x_i)&=&3(x_2-x_1)\psi_5^0-3(x_2-x_1)x_3\psi_5^-+3(x_2-x_1)(1-2x_2-2x_1)\psi_5^+,\nonumber\\
A_4(x_i)&=&-3(1-x_3)\psi_5^0-3(2x_1x_2-(x_1+x_2))\psi_5^--3(x_1+x_2)(1-2x_1-2x_2)\psi_5^+,\nonumber\\
V_5(x_i)&=&6x_3(x_1-x_2)\phi_5^-,\hspace{2cm}A_5(x_i)=-6x_3[\phi_5^0+\phi_5^+(1-2x_3)],\nonumber\\
T_4(x_i)&=&-\frac32(x_1-x_2)(\xi_5^{'0}+\xi_5^0)-\frac32x_3(x_2-x_1)(\xi_5^{'-}+\xi_5^-)\nonumber\\
&&-\frac32(x_1-x_2)(1-2x_1-2x_2)(\xi_5^{'+}+\xi_5^+),\nonumber\\
T_5(x_i)&=&6x_3[t_5^0+t_5^-(x_1-x_2)+t_5^+(1-2x_3)],\nonumber\\
T_8(x_i)&=&-\frac32(x_1-x_2)(\xi_5^0-\xi_5^{'0})-\frac32x_3(x_2-x_1)(\xi_5^--\xi_5^{'-})\nonumber\\
&&-\frac32(x_1-x_2)(1-2x_1-2x_2)(\xi_5^+-\xi_5^{'+}).
\end{eqnarray}
Finally twist-$6$ distribution amplitudes are
\begin{eqnarray}
V_6(x_i)&=&2\phi_6^-(x_1-x_2),\hspace{2.5cm}A_6(x_i)=-2[\phi_6^0+\phi_6^+(1-3x_3)],\nonumber\\
T_6(x_i)&=&2[t_6^0+t_6^-(x_1-x_2)+t_6^+(1-3x_3)].
\end{eqnarray}

\section{Summary}\label{sec:summary}

We present the improved LCDAs of the $\Lambda$ baryon up to twist $6$. Our calculations are based on the conformal symmetry of the massless QCD Lagrangian. Using the relations from the isospin property $I=0$ of the baryon, the number of the independent LCDAs is reduced to $14$. The LCDAs defined with chiral-field representations are expanded according to the conformal spin to the next-to-leading order. The next-to-leading order corrections of the LCDAs come from the expansion of the nonlocal three-quark operator matrix element at zero point to the second order. In consideration of the general Lorentz decomposition, the three matrix elements of the local three-quark operator can be parameterized with $24$ variables.  With the constrains from the equations of motion, the number of the independent parameters is reduced to $10$.

The independent $10$ parameters can be related to the coupling constants defined by the matrix elements of different Lorentz structures between the vacuum and the baryon state. Then the conformal expansion parameters are expressed by these coupling constants. Finally the QCD sum rules method is used to estimate these constants. The explicit expressions of the LCDAs of the $\Lambda$ baryon are presented as the main results of this paper.

\acknowledgments  This work was supported in part by the National
Natural Science Foundation of China under Contracts No. 11475257, No.11105222, and No.11275268.

\end{document}